\documentclass{moriond}

\usepackage{amssymb}

\def\Journal#1#2#3#4{{#1} {\bf #2}, #3 (#4)}

\def\be{\begin{equation}}
\def\ee{\end{equation}}
\def\bea{\begin{eqnarray}}
\def\eea{\end{eqnarray}}

\newcommand{\Photo}{}

\begin{document}
\vspace*{4cm}
\title{Electroweak results in two-photon collisions}

\author{ C. Caillol, on behalf of the ATLAS and CMS Collaborations}

\address{CERN, Switzerland}

\maketitle\abstracts{
The LHC is also a high-energy photon collider. The ATLAS and CMS experiments are exploring a wide diphoton energy range with various approaches in heavy-ion and proton collisions to precisely measure quantum electrodynamics processes and constrain the existence of physics beyond the standard model.}

\section{Introduction}

Since the beginning of its operation, the LHC has produced an impressive amount of significant results based on very high energy proton-proton collisions, leading, among others, to the discovery of a new scalar boson and the exclusion of a large phase space of supersymmetry. The heavy-ion collision programme has been as successful, with hundreds of publications. Experiments are now pushed to explore their broad range of accessible physics processes to potentially uncover the presence of physics beyond the standard model (BSM physics). In this context, the physics programme related to photon-induced processes has intensified in recent years. Indeed, the LHC can also be considered as the highest energy photon collider.

Photon-induced processes can occur when two charged objects, like protons or ions, pass each other at relativistic velocities, generating intense electromagnetic fields. The cross section of photon-induced processes is proportional to $Z^4$, where Z is the atomic number. This leads to a large enhancement of the cross section in Pb-Pb runs, where $Z=82$, compared to proton-proton (pp) runs, where $Z=1$. However, this effect competes with the much larger luminosity available in pp runs. 

\begin{figure}
\begin{minipage}{1.0\linewidth}
\centerline{\includegraphics[width=0.7\linewidth]{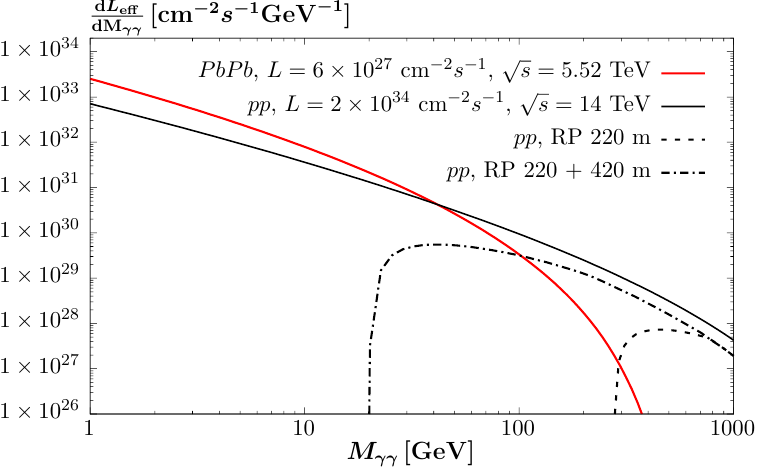}}
\end{minipage}
\caption[]{Effective $\gamma\gamma$ luminosity vs. photon-fusion mass in ultraperipheral Pb-Pb and pp collisions at the LHC.\cite{th}}
\label{fig:th}
\end{figure}

The effective luminosity of $\gamma\gamma$ processes can be compared for Pb-Pb runs and pp runs as a function of the mass, as shown in Fig. \ref{fig:th} assuming an instantaneous luminosity L of $6\times10^{27}\textrm{cm}^{-2}\textrm{s}^{-1}$ and a center-of mass energy $\sqrt{s}=5.52$ TeV for Pb-Pb collisions, and $2\times10^{34}\textrm{cm}^{-2}\textrm{s}^{-1}$ and 14 TeV for pp collisions \cite{th}. One can identify 3 different mass regions:
\begin{itemize}
\item Low diphoton masses ($m_{\gamma\gamma}\lessapprox 30$ GeV). The effective luminosity is highest in Pb-Pb runs and ultraperipheral events with objects with low transverse momentum ($p_T$) can be selected online because of the very clean environment.
\item Intermediate diphoton masses ($30\lessapprox m_{\gamma\gamma}\lessapprox 350$ GeV). The effective luminosity in pp runs starts dominating. The environment in pp collisions is very busy with a high number of tracks arising from pileup collisions, which constitutes an experimental challenge to identify ultraperipheral collisions.
\item High diphoton masses ($m_{\gamma\gamma}\gtrapprox350$ GeV). Forward detectors (PPS for CMS and AFP for ATLAS) can tag diffracted protons, and matching the kinematics of the forward and central systems can be used as an extra handle to separate photon-induced events from the background.
\end{itemize}

In the rest of this note, we will cover searches and measurements by the ATLAS and CMS Collaborations in these three different regimes, focussing on the most recent results.

\section{The low energy regime: Pb-Pb ultraperipheral collisions}

Thanks the the $Z^4$ enhancement, Pb-Pb ultraperipheral collisions are the best environment at the LHC to study photon-induced processes at low diphoton masses. As photon-induced processes are pure quantum electrodynamics processes, predicted with a high level of accuracy, their study is a particularly powerful test of the SM. Precise measurements of the $\gamma\gamma\rightarrow ee$ and $\gamma\gamma\rightarrow\mu\mu$ processes have been performed by the ATLAS and CMS Collaborations, including differential measurements with respect to various observables, and compared to the predictions of different generators. The $\gamma\gamma\rightarrow\gamma\gamma$ process, also known as "light-by-light scattering", was also studied and observed by both experiments.

\begin{figure}
\begin{minipage}{0.45\linewidth}
\centerline{\includegraphics[width=0.9\linewidth]{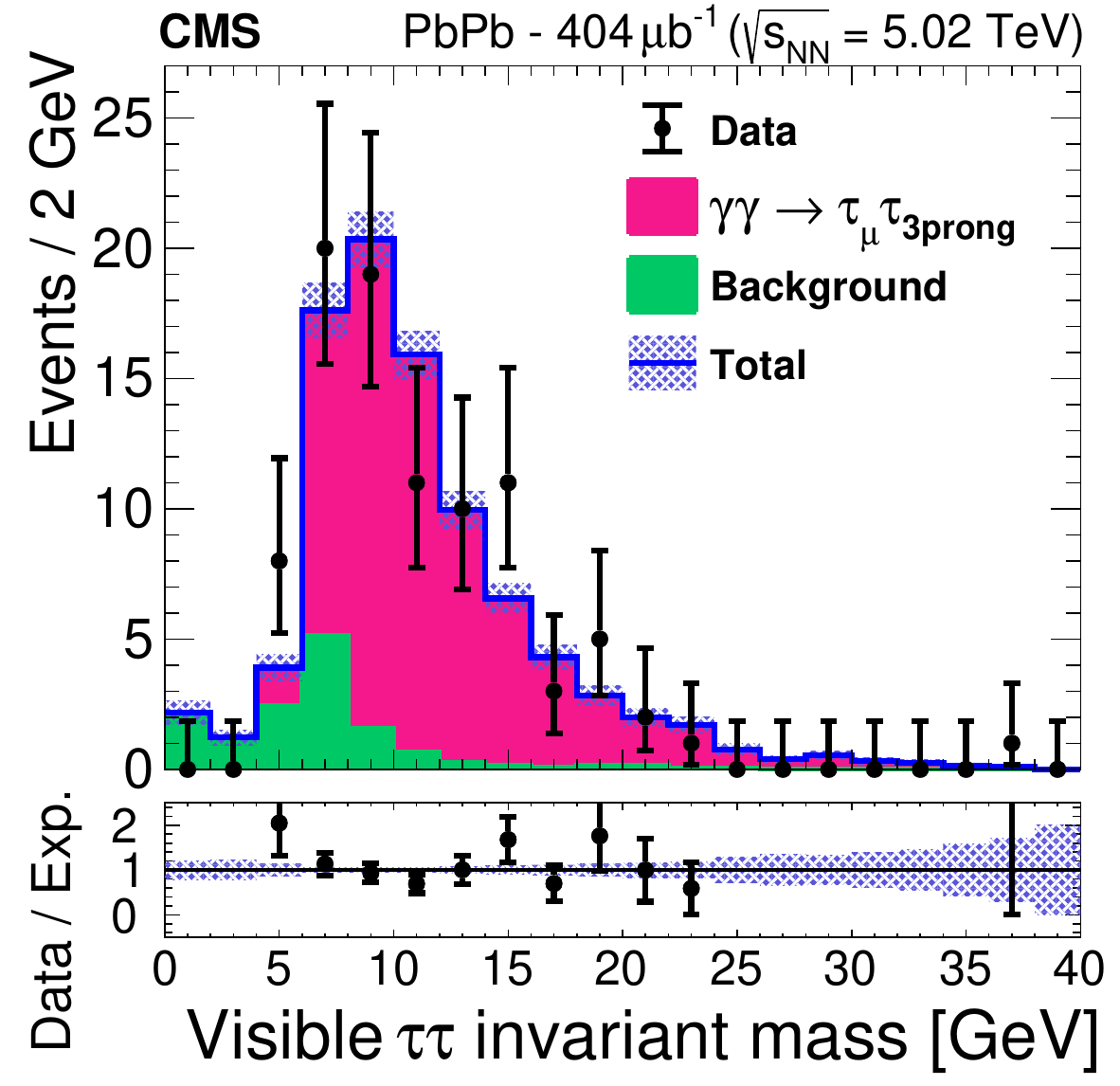}}
\end{minipage}
\hfill
\begin{minipage}{0.45\linewidth}
\centerline{\includegraphics[width=0.9\linewidth]{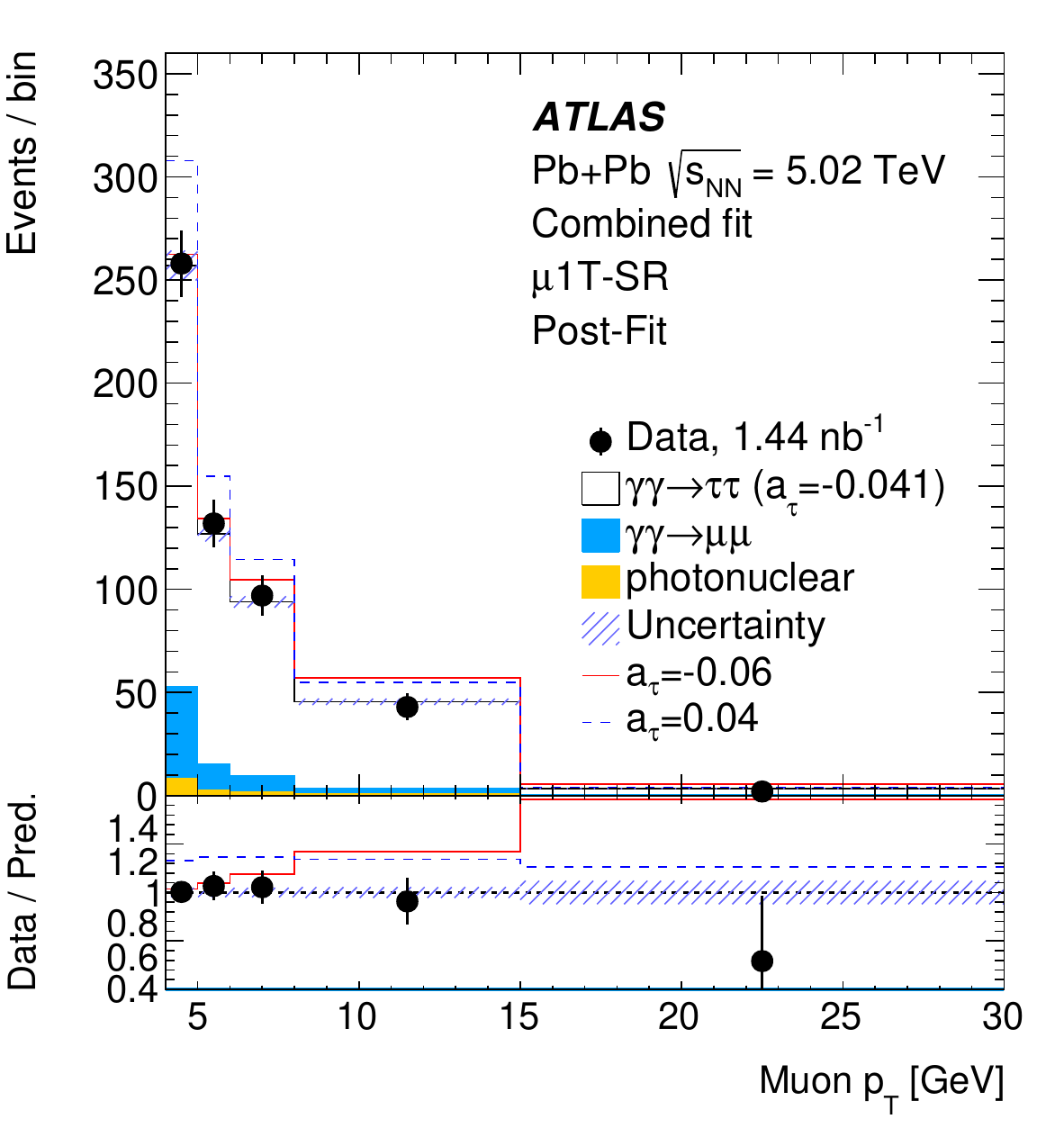}}
\end{minipage}
\caption[]{Left: Visible di-$\tau$ invariant mass distribution for events collected by the CMS experiment in 2015 in the final state with one muon and one hadronically decaying $\tau$ lepton with 3 tracks. Right:  Distribution of the muon $p_T$ for events collected by the ATLAS experiment in 2018 in the final state of one muon and one hadronically decaying $\tau$ lepton with 1 track. Predicted distributions for BSM values of $a_\tau$ are shown with the full red and dashed blue lines.}
\label{fig:pbpb}
\end{figure}

The ATLAS and CMS experiments both observed the $\gamma\gamma\rightarrow\tau\tau$ process in Pb-Pb collisions using data collected during the Run-2 of the LHC, with a significance well above the 5 standard deviations level \cite{ATLASHI,CMSHI}. While the CMS analysis relies on $0.4\textrm{nb}^{-1}$ of Pb-Pb collisions collected in 2015 and on the decay channel with one muon and one hadronically decaying $\tau$ lepton with 3 tracks, the ATLAS result is based on $1.4\textrm{nb}^{-1}$ of Pb-Pb collisions collected in 2018 and on the decay channels with one muon and one hadronically decaying $\tau$ lepton with 3 tracks, one electron and one muon, and one muon and one hadronically decaying $\tau$ lepton with 1 track. As shown in Fig. \ref{fig:pbpb} (left), the visible ditau invariant mass range, calculated using the visible decay products of the tau leptons, is typically between 5 and 25 GeV for reconstructed signal events. It can be noted that the signal-to-background ratio is very high because of the clean environment in Pb-Pb ultraperipheral collisions.

Information about the  $\tau$ anomalous magnetic moment ($\tau$ g-2) can be extracted from the $\gamma\gamma\rightarrow\tau\tau$ process because of the presence of two $\gamma\tau\tau$ vertices. A modification of the $\tau$ g-2 from the SM prediction would modify the cross section of the signal process as well as the $p_T$ distribution of the decay products, as shown in Fig. \ref{fig:pbpb} (right), where $a_{\tau}=(g-2)/2$. The CMS analysis makes use of cross section information, while the ATLAS analysis relies on the $\tau$ $p_T$ distribution to set constraints on $a_{\tau}$. The 68\% confidence level interval measured by the ATLAS Collaboration is slightly wider than the interval measured twenty years ago by the DELPHI experiment at LEP.

\section{The intermediate energy regime: pp collisions with track counting}

Accessing a higher diphoton mass regime can be useful when probing the photon-induced production of heavier particles, e.g. $\gamma\gamma\rightarrow WW$, or when BSM effects increase with the mass, as it is the case for $a_{\tau}$ measurements in $\gamma\gamma\rightarrow\tau\tau$ events. The main handle to identify ultraperipheral collisions is the absence of additional tracks from the hard interaction. However, all events, including signal events, are swamped with hundreds of tracks coming from additional pp interactions, called pileup interactions. To identify signal events, the analyses count the number of tracks in very narrow windows in the direction of the beamline, centered around the vertex made from the final state particles ($\tau$ lepton or W boson visible decay products). 

\begin{figure}
\begin{minipage}{0.45\linewidth}
\centerline{\includegraphics[width=1.0\linewidth]{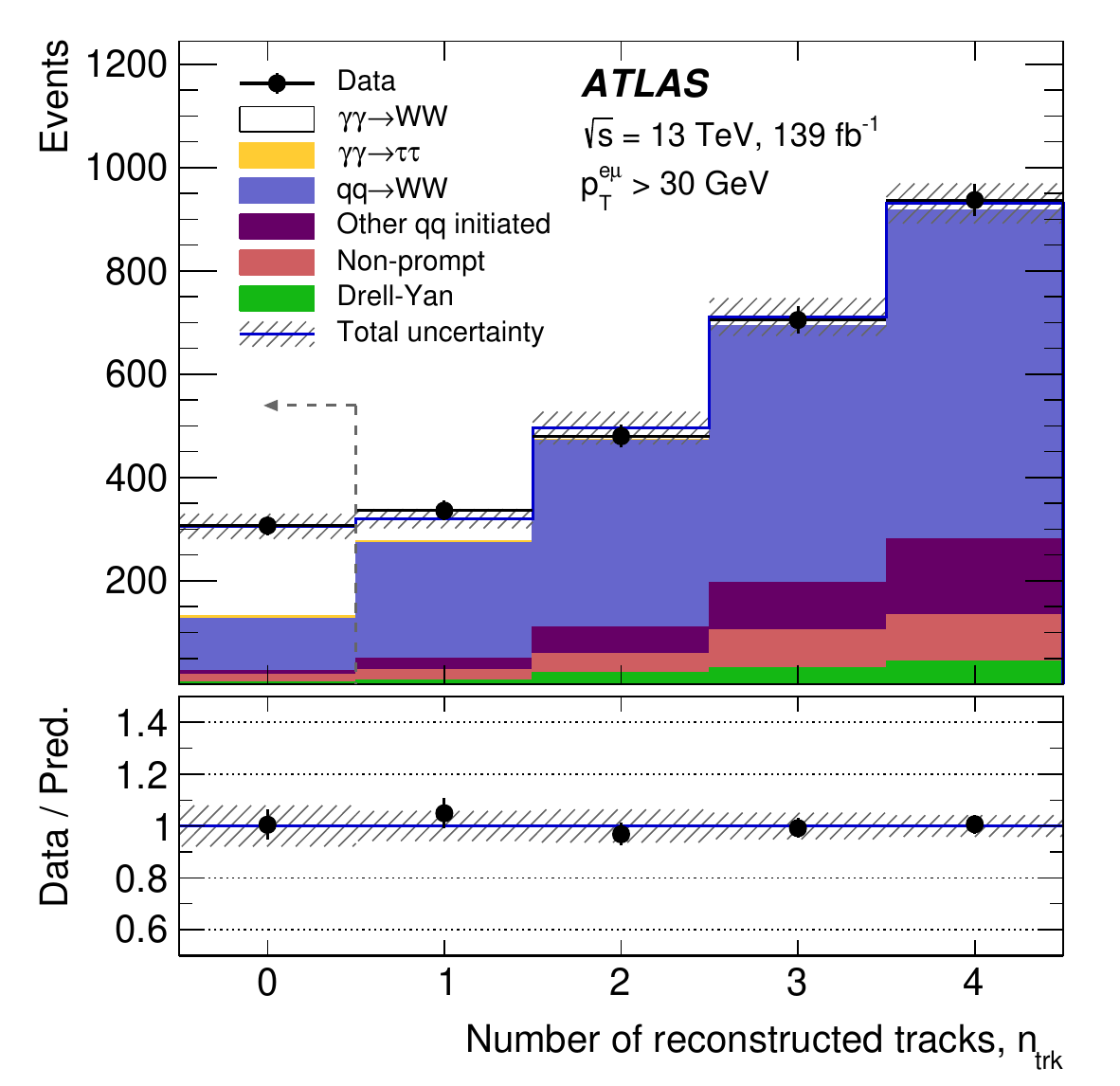}}
\end{minipage}
\hfill
\begin{minipage}{0.45\linewidth}
\centerline{\includegraphics[width=1.0\linewidth]{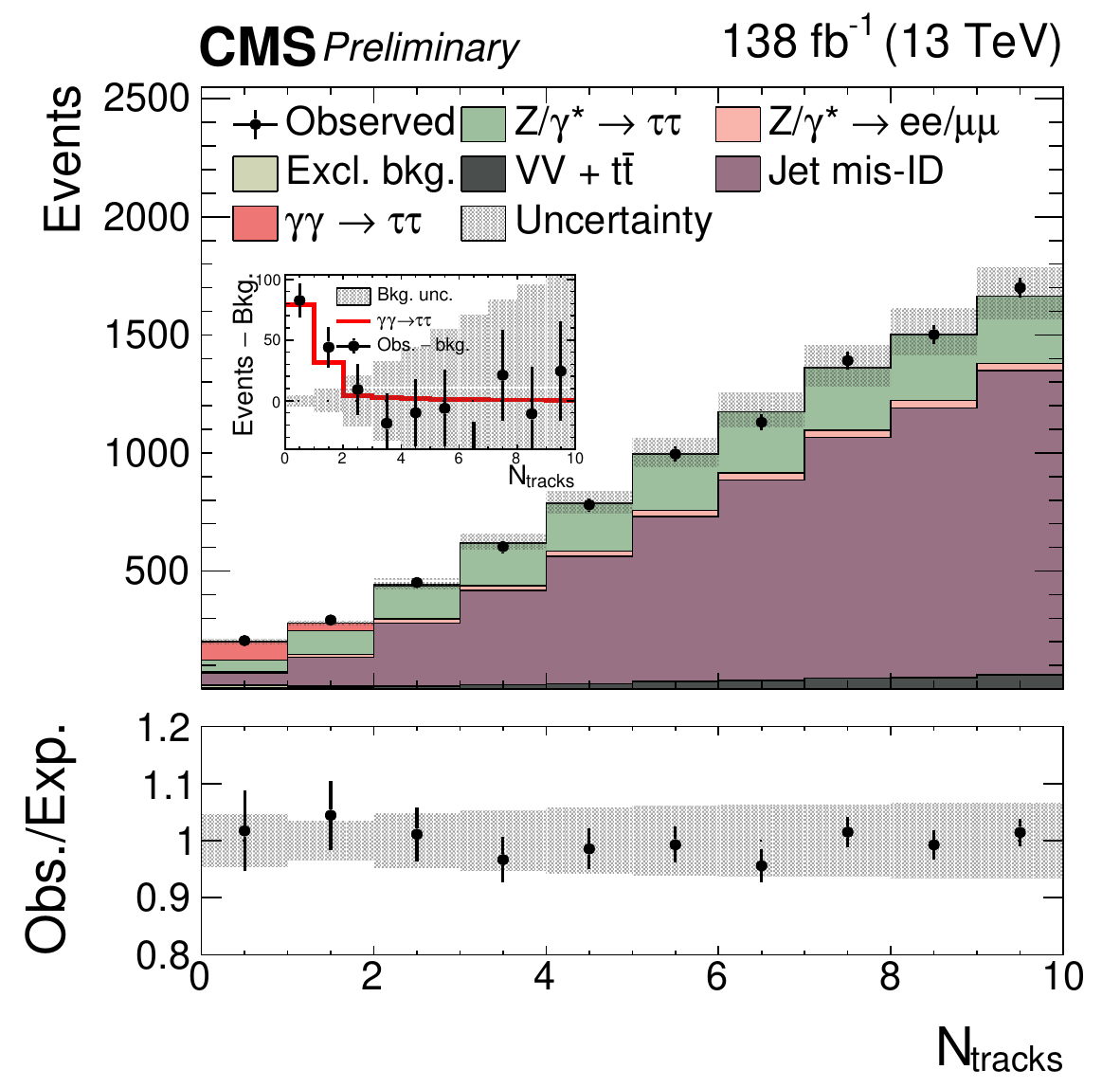}}
\end{minipage}
\caption[]{Distributions of the number of additional tracks for $WW$ (left) and $\tau\tau$ (right) events. The signal is visible in the first bins. }
\label{fig:pp}
\end{figure}

Using Run-2 data, the ATLAS and CMS experiments announced the first observations in pp collisions,of $\gamma\gamma\rightarrow WW$ and $\gamma\gamma\rightarrow\tau\tau$, respectively \cite{ATLASWW,CMSTT}. The number of tracks with $p_T>0.5$ GeV and $|\eta|<2.5$ in windows of 0.1 or 0.2 cm width, excluding tracks associated with the W bosons and $\tau$ leptons, are required to be 0 or 1. Signal and background simulations need to be corrected to accurately model the track multiplicity. These corrections are obtained from a dimuon control region, comparing between data and simulation the track multiplicity far from the dimuon vertex to access pileup tracks, and close to the dimuon vertex to access hard scattering tracks. The signal is modelled using an elastic simulation, and the fraction of dissociative contributions is evaluated from dimuon exclusive events, rescaling the elastic simulation to describe the tails of the dimuon mass distribution observed in data in events without additional tracks. The signals are clearly visible as excesses of events over the background prediction for events with few additional tracks, as shown in Fig. \ref{fig:pp}.

\begin{figure}
\begin{minipage}{0.45\linewidth}
\centerline{\includegraphics[width=1.0\linewidth]{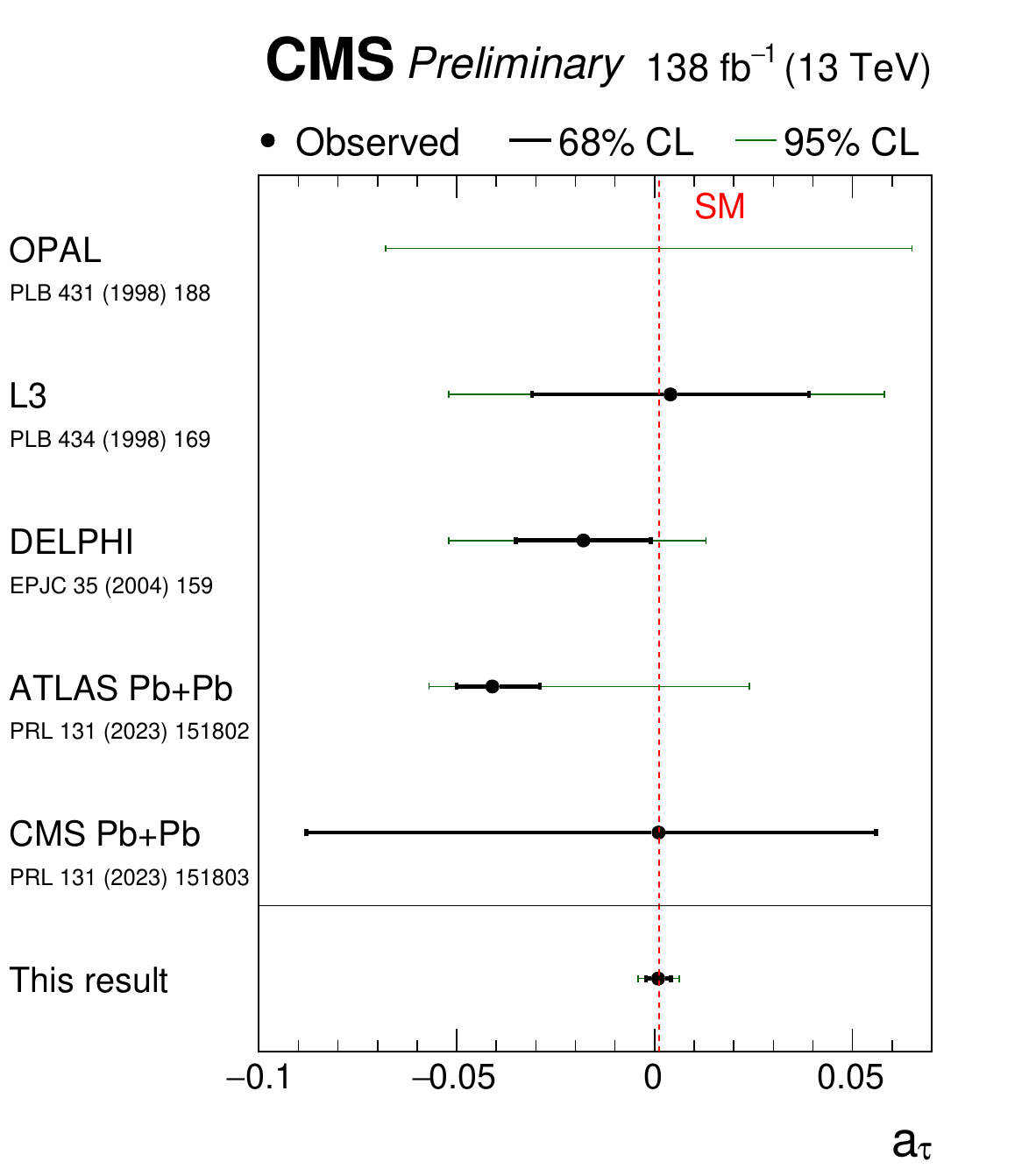}}
\end{minipage}
\hfill
\begin{minipage}{0.45\linewidth}
\centerline{\includegraphics[width=1.0\linewidth]{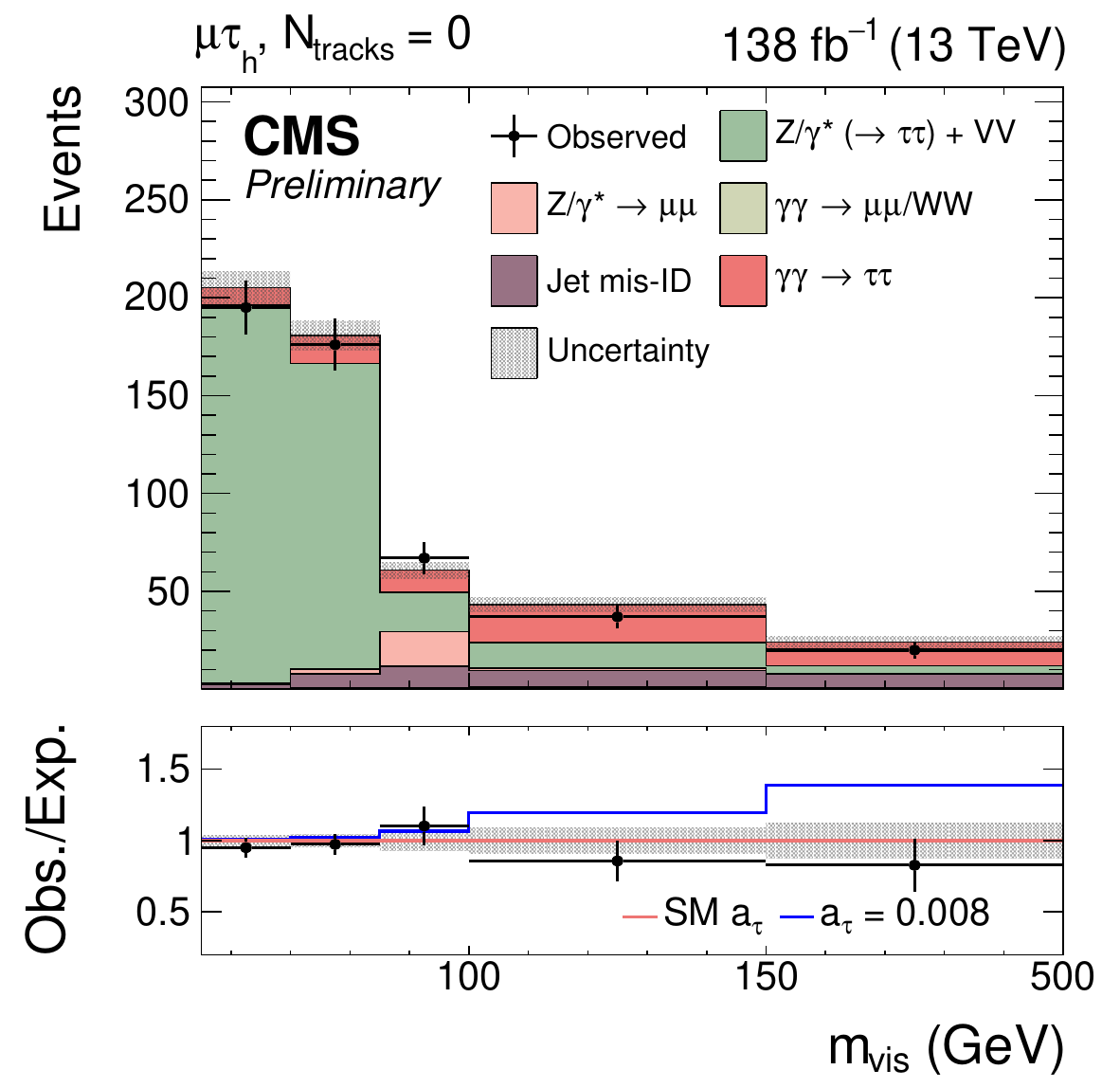}}
\end{minipage}
\caption[]{Left: Constraints on $a_\tau$ set by the recent CMS analysis in pp collisions, compared with past constraints from other measurements or experiments. Right: Distributions of the visible di-tau invariant mass in the $\mu\tau_h$ final state with no additional track. The signal contribution for an illustrative value of $a_\tau$ is shown with the blue line in the ratio panel.}
\label{fig:g2pp}
\end{figure}

The CMS Collaboration used the di-tau visible invariant mass distributions of $\gamma\gamma\rightarrow\tau\tau$ events to set the tightest constraints on $a_{\tau}$, improving the LEP results \cite{DELPHI} by about one order of magnitude, as shown in Fig. \ref{fig:g2pp} (left). The largest sensitivity compared to measurements in heavy-ion collisions comes from the increase of BSM effects with the mass of the di-$\tau$ system, which is effectively much larger in the events accessed by the pp analysis. As shown in Fig. \ref{fig:g2pp} (right) for the $\mu\tau_h$ final state with no additional track, the signal enhancement with $a_{\tau}$ grows with the di-tau visible invariant mass. Constraints were set using an SMEFT approach, where BSM contributions to $a_{\tau}$ from heavy new degrees of freedom can be parameterized using Wilson coefficients. A similar approach is used to set constraints on the electric dipole moment of the $\tau$ lepton. The results are extracted by performing negative log-likelihood scans as a function of $a_{\tau}$, modifying the visible di-tau invariant mass distributions of the signal accordingly using matrix element reweighting. 

\section{The high energy regime: pp collisions with proton tagging in forward detectors}

When the diphoton mass is high enough ($m_{\gamma\gamma} \gtrapprox 350$ GeV), both diffracted protons are in the acceptance of the forward proton detectors (PPS for CMS, AFP for ATLAS) and the reconstructed protons can be used to tag events. This approach is particularly useful when the interaction vertex cannot be reconstructed such that the track multiplicity cannot be determined, e.g. in the case of di-photon production, or when the mass of the particles produced is very high, e.g. in the context of searches for new heavy BSM particles.

Recent searches by the CMS and ATLAS Collaborations for photon-induced events involving the forward proton detectors include searches in the final state of two photons, in the context of light-by-light scattering mediated by axion-like particles (ALPs)\cite{ATLASALP,CMSALP}. The ATLAS analysis \cite{ATLASALP} selects diphoton events if there is kinematic matching with a proton in at least one side of AFP, and extracts the results from an unbinned maximum likelihood fit to the diphoton invariant mass distribution. The largest local significance is 2.5 standard deviations. The CMS analysis \cite{CMSALP} matches the mass and rapidity of the pp and $\gamma\gamma$ systems within two standard deviations. This selects 1 data event, for a background expectation of 1.1 event, leading to constraints on ALPs with mass between 0.5 and 2 TeV. The same events are also used to constrain anomalous quartic gauge coupling parameters from the high mass exclusive diphoton production.

\section{Conclusion}

The photon-induced physics programme of the LHC is blooming, pushing further the range of physics accessible with the ATLAS and CMS detectors. While $\gamma\gamma$ collisions were initially studied essentially in heavy-ion peripheral collisions as precision tests of quantum electrodynamics, ATLAS and CMS have now demonstrated that the amazing tracking capabilities of the detectors could be used to separate photon-induced interaction vertices from pileup vertices in proton-proton collisions and observed processes as elusive as $\gamma\gamma\rightarrow WW$ and $\gamma\gamma\rightarrow\tau\tau$. The higher invariant mass of the $\tau\tau$ system compared to that available in heavy-ion collisions allowed the CMS experiment to set the tightest constraints on $\tau$ g-2, improving constraints from LEP by almost an order of magnitude. Forward proton detectors play a significant role in photon-induced searches at high diphoton masses, allowing the kinematics of the central and forward systems to be matched to reduce the backgrounds significantly. All these recent developments open the path for an exciting photon-collision programme at the LHC, pushing even further the physics reach to directions that had not been initially expected.

\end{document}